 \documentclass[final,11pt,1p]{elsarticle}

\usepackage{amssymb}

\begin{document}

\begin{frontmatter}

\title{Applied-Field Magnetoplasmadynamic Thrusters for Deep Space Exploration}

\author{Matthew Han}
\author{Hannah Rana\corref{cor1}}
\address{Center for Astrophysics $|$ Harvard \& Smithsonian, 60 Garden St, Cambridge, MA 02138, USA}

\cortext[cor1]{Corresponding author}

\ead{hannah.rana@cfa.harvard.edu}

\begin{abstract}
Recent advancements in the development of Applied-Field Magnetoplasmadynamic thrusters (AF-MPDTs) present themselves to be an increasingly promising propulsion technology for deep space exploration missions. Various entities, ranging from state-sponsored institutions to privately-owned startups, have developed AF-MPDTs across a wide range of power levels. Current developments in superconducting technologies, namely High-Temperature Superconducting (HTS) coils such as REBCO, have enabled research into the integration of HTS coils into the applied-field module to generate MPD thrust. Developments in space cryocoolers have opened the doors for HTS use within a spaceflight design of an AF-MPDT, where the applied-field module is at 40\,K. A TRL of 4-5 has been reached by some AF-MPDT prototypes; venturing beyond this will require higher cooling power space cryocoolers to be developed in parallel and appropriately integrated into the thruster. Moreover, bespoke thermal control is required to maintain the thruster's extreme temperature gradient where the cryocooled HTS are in close proximity to the thruster cathode. More effective space power supply units with higher power generation is a further limitation to reaching TRL 9. This review examines the underlying principles behind AF-MPDT propulsion and the recent global developments in AF-MPDT technology, with an in-depth analysis and critical discussion on the spaceflight components necessary to permit AF-MPDTs to become a widely-adopted spaceflight-ready propulsion technology.
\end{abstract}


\end{frontmatter}


\section{Introduction}\label{sec1}
Chemical propulsion systems have been the traditional choice to launch a rocket out of the Earth’s atmosphere. Chemical propulsion utilizes combustion, an exothermic reaction, between a fuel and an oxidizer to accelerate the exhaust, hence creating thrust. Therefore, the exhaust velocity is limited to the chemical properties of the propellants, since they have set combustion temperatures. This puts a relatively low cap on the specific impulse of the rocket, as specific impulse is directly related to exhaust velocity. The maximum theoretical specific impulse of a chemical propulsion rocket is about 500 ISP, which is the unit of specific impulse. However, the simplicity of combustion allows the mass flow rate in chemical propulsion to be scaled up to extremely high levels, which creates the high thrust needed to escape Earth’s gravity \cite{mishra2017}.

Electric propulsion, on the other hand, is a relatively under-developed technology that could be useful for in-space propulsion. Electric propulsion utilizes a magnetic field to accelerate charged ions in order to create thrust. Current electric propulsion technologies have achieved specific impulses up to 5000 ISP and can theoretically achieve specific impulses up to 15,000 ISP. The downside to electric propulsion is its extremely low mass flow rate and the challenges with scaling it. However, electric propulsion’s extremely high efficiency makes it useful for in-space propulsion, away from the effects of Earth’s gravity \cite{brophy1990}.\\
\\

One emerging type of EP is the AF-MPDT. this type of propulsion enables travelling to further distances and is seen as a promising enabler for missions to Mars. This paper will lay out the different types of electric propulsion thrusters and focus on the state of the art and developments of AF-MPDTs.

\section{Background Theory}
\subsection{Types of Electric Propulsion}
\subsubsection{Electrothermal Propulsion}
Electrothermal propulsion utilizes electric power to heat a propellant in order to increase its efficiency. The heating could be done by simply heating the propellant through an electric arc or by using a catalyst. These systems are usually more reliable than the other electric propulsion types but offer a lower specific impulse, usually not surpassing 1000 seconds. Types of electrothermal propulsion include the arcjet and the resistojet \cite{jahn1968}.

\subsubsection{Electrostatic Propulsion}
Electrostatic propulsion uses electric fields to accelerate ionized propellant in order to produce thrust. The most common propellants are Cesium, Mercury, Xenon, and Argon. The ions are accelerated using an electric field created by a series of grids before becoming neutralized by an electron stream in order to prevent a charge build up. Electrostatic propulsion produces an extremely high specific impulse, around 3000-5000 seconds. However, these thrusters can only provide a very low thrust. Therefore, electrostatic propulsion is mainly used for orbital transfers because of its superior propellant efficiency and the relative unimportance of time needed to complete in-orbit transfers. These are the most commonly used electric propulsion technologies in space, boasting the most flight heritage out of the various electric propulsion systems. Types of electrostatic propulsion include the gridded ion thruster and the Hall effect thruster \cite{jahn1968}.\\

\subsubsection{Electromagnetic Propulsion}
Electromagnetic propulsion creates thrust by utilizing the Lorentz force to accelerate a plasma. This is done by using magnetic fields to interact with electric currents in the plasma, accelerating it to create thrust. This technique allows for control over the thrust level and direction \cite{myers1993}. Compared to electrostatic propulsion, electromagnetic propulsion produces thrust levels 100 times larger than electrostatic propulsion while maintaining high efficiency. However, electromagnetic propulsion thrusters are not as common for orbital transfers because of the complexity of the engine. One aspect that makes the engines complex is that large thermal gradients are required between various subcomponents that are in close proximity to one another. For example, in the case of AF-MPDTs, superconducting coils are used to generate thrust as the plasma passes through their electromagnetic field. These coils require cooling to 40-60\,K, whilst residing within the thruster design in very close proximity to the thruster nozzle, which reaches greater than 1000K in temperature. Thus far, this thermal control related challenge has been the largest obstacle to overcome in bringing AF-MPDTs to the space market. Types of electromagnetic propulsion thrusters include the magnetoplasmadynamic and the pulsed plasma thrusters \cite{jahn1968}.

\subsection{AF-MPDT Principles}
The AF-MPDT is an electromagnetic propulsion thruster that utilizes electrical currents and magnetic fields to accelerate an ionized propellant. MPDTs can be classified into two main types based on how the magnetic field is generated: self-field and applied-field MPDTs. In self-field MPDTs, the magnetic field is generated by the current passing through the plasma itself. This current, driven by a voltage difference between the cathode and anode, induces an azimuthal magnetic field around the axis of the thruster. In applied-field MPDTS, an electromagnet or permanent magnet is placed around the thruster, creating an axial magnetic field that is significantly stronger than the self-induced magnetic field \cite{andrenucci2010}. Although this review focuses on the AF-MPDT, the operating principles for the self-field MPDT will be mentioned as it is still present in the AF-MPDT.

\begin{figure} [h]
    \centering
    \includegraphics[width=0.6\linewidth]{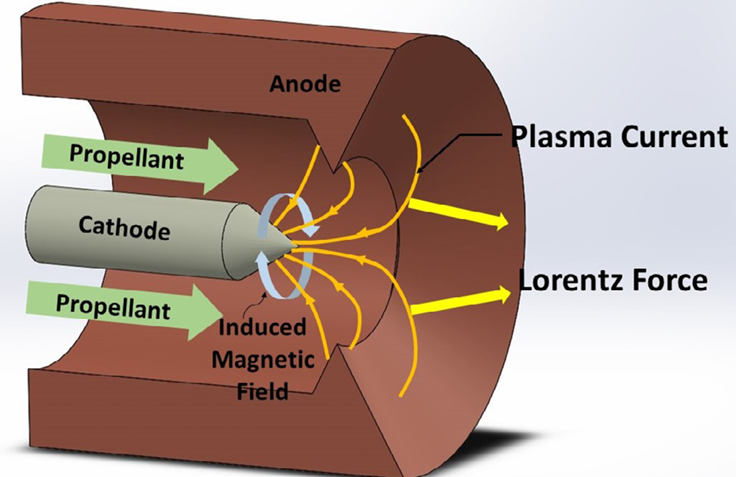}
    \caption{A simplified schematic showing the cross-section of an SF-MPDT \cite{wang2021}.}
    \label{simplified-sfmpdt}
\end{figure}

\subsubsection{Self Magnetic Field}
In self-field MPDTs, the magnetic field is generated by the current passing through the plasma itself. This current, driven by a voltage difference between the cathode and anode, induces an azimuthal magnetic field around the axis of the thruster \cite{kodys2005}. The relationship governing this induced magnetic field is given by Ampere's Law \cite{sanchez2014}:
\begin{equation}
\nabla \times \mathbf{B} = \mu_0 \mathbf{j}
\end{equation}

In integral form, this can be expressed as:

\begin{equation}
\oint_\Gamma \frac{\mathbf{B}}{\mu_0} \cdot d\mathbf{l} = \int_\Sigma \mathbf{j} \cdot d\mathbf{A}
\end{equation}\\

This law implies that the circulation of a magnetic field around a closed loop is proportional to the current passing through the surface enclosed by the loop. For any radial cross-section of the thruster, this means that the azimuthal magnetic field \(B_\theta\) is directly related to the current density \(j\).

\subsubsection{The Lorentz Force}
The Lorentz force is the primary mechanism for plasma acceleration in MPDTs \cite{andrenucci2010}. It acts on the ions and electrons in the plasma, driving the acceleration process. The momentum equation for a plasma can be written separately for ions and electrons \cite{sanchez2014}:

For ions:
\begin{equation}
m_i n \frac{D\mathbf{u}_i}{Dt} = n e (\mathbf{E} + \mathbf{u}_i \times \mathbf{B}) - \nabla p_i + \mathbf{P}_{ie}
\end{equation}

For electrons:
\begin{equation}
0 = -n e (\mathbf{E} + \mathbf{u}_e \times \mathbf{B}) - \nabla p_e + \mathbf{P}_{ei}
\end{equation} 

Here, \(m_i\) is the ion mass, \(n\) is the plasma density, \(e\) is the charge of an electron, \(\mathbf{E}\) is the electric field, \(\mathbf{B}\) is the magnetic field, \(\mathbf{u}_i\) and \(\mathbf{u}_e\) are the ion and electron velocities, \(p_i\) and \(p_e\) are the ion and electron pressures, and \(\mathbf{P}_{ie}\) and \(\mathbf{P}_{ei}\) are the momentum exchange terms between ions and electrons.

When you look at the overall effect on the plasma, the combined movement of ions and electrons can be simplified down to:

\begin{equation}
\rho \frac{D\mathbf{u}}{Dt} = \mathbf{j} \times \mathbf{B} - \nabla p
\end{equation}

where \(\rho\) is the mass density of the plasma, \(\mathbf{u}\) is the bulk velocity of the plasma, and \(\mathbf{j}\) is the current density.

The Lorentz force per unit volume \(\mathbf{j} \times \mathbf{B}\) drives the plasma acceleration, while the pressure gradient \(\nabla p\) acts to resist this acceleration. In efficient MPDT operation, the electromagnetic term \(\mathbf{j} \times \mathbf{B}\) dominates over the pressure term.

\subsubsection{Performance and Efficiency}

To evaluate the performance of an MPDT, the mechanical work done by the Lorentz force on the plasma must be examined, which can be done by looking at how the kinetic energy of the ions increases \cite{sanchez2014}. The kinetic energy equation for ions can be derived by taking the scalar product of the momentum equation with the ion velocity \(\mathbf{u}_i\) \cite{sanchez2014}:

\begin{equation}
\frac{D}{Dt} \left( \frac{1}{2} \rho u_i^2 \right) = (\mathbf{j} \times \mathbf{B}) \cdot \mathbf{u}_i - \nabla p \cdot \mathbf{u}_i
\end{equation}

This equation indicates that the increase in kinetic energy of the plasma mainly comes from the work done by the Lorentz force with a smaller contribution by the pressure gradient. 

For electrons, the generalized Ohm's law can be derived:

\begin{equation}
\mathbf{E} = \frac{\mathbf{j}}{\sigma} - \frac{\nabla p_e}{ne} + \frac{\mathbf{j} \times \mathbf{B}}{ne} - \mathbf{u}_i \times \mathbf{B}
\end{equation}

where \(\sigma\) is the electrical conductivity. The term \(\frac{\mathbf{j}}{\sigma}\) represents Ohmic heating, which is a source of inefficiency as it converts useful energy into heat\cite{andrenucci2010} \cite{sanchez2014}.

\subsubsection{Applied-Field MPDTs}

\begin{figure}[h]
    \centering
    \includegraphics[width=0.8\linewidth]{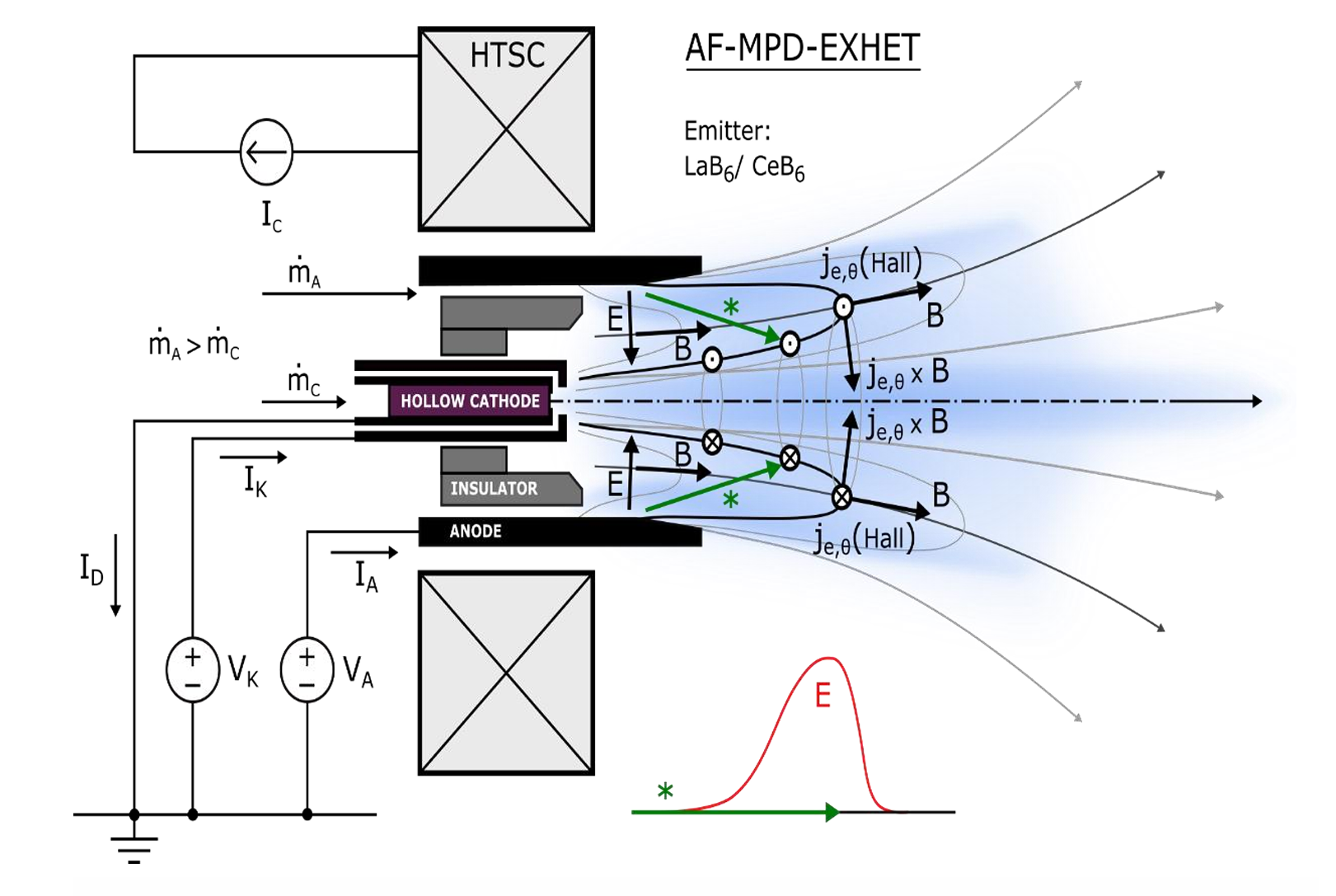}
    \caption{A schematic showing the principles of the AF-MPDT thrust mechanism \cite{boxberger2017}.}
    \label{nss-image}
\end{figure}

The fundamental difference between the AF-MPDT and the SF-MPDT is the external magnetic field that enhances the performance of the AF-MPDT. The external magnetic field is primarily axial, but includes a small radial component. This applied field dwarfs the self-induced magnetic field, which is generated by the flow of the current through the plasma. Including the self-field acceleration, four main forces accelerate the propellant in an AF-MPDT\cite{kodys2005}. In Figure \ref{nss-image} a schematic of the operation of an AF-MPDT is shown.

\textbf{Self-field Acceleration}\\
Self-field acceleration relies on the Lorentz force, which arises from the interaction between the azimuthally induced magnetic field and the radial current \cite{gilland1991}. The Lorentz force can be noted as:
\begin{equation}
F_{z} = j_{r} \times B_{\theta}
\end{equation}

The Lorentz Force produces a force in two directions. First, it pushes the plasma axially outwards, directly contributing to the thrust. Secondly, it pushes the plasma radially inward, called the pinching force. The pinching force creates a pressure imbalance, indirectly contributing to the thrust. The force generated by this mechanism can be expressed as \( j_r B_\theta \) \cite{choueiri1998}. \\

\textbf{Swirl Acceleration}\\
Another component of the Lorentz force causes the plasma to swirl around the cathode. An expanding nozzle converts this swirling energy into forward thrust. Increasing either the current or magnetic field strength results in more thrust being generated by swirl acceleration. Swirl Acceleration is often observed as the dominant thrust producer in AF-MPDTs\cite{cann1967}. \\

\textbf{Hall Acceleration}\\
A circular current can be created according to Ohm’s Law at high magnetic field strengths and low mass flow rates. Similar to self-field acceleration, pinching and blowing forces are created. However, whether these forces positively or negatively contribute to the thrust is unknown. Either way, it is most likely a tiny contribution to the total thrust\cite{sasoh1994}. \\

\textbf{Gas Dynamic Acceleration}\\
Like chemical propulsion, gas dynamic acceleration relies on heating and expanding the plasma through the nozzle. Scaling this force is driven primarily by the mass flow rate of the propellant, with less dependence on current and magnetic fields. Gas dynamic acceleration accounts for a significant portion of the thrust\cite{kodys2005}. 

\subsubsection{Onset Phenomena}

The onset phenomena is a critical limiting effect in the steady-state operation of AF-MPDTs. This phenomenon is a plasma instability that occurs when the current exceeds a specific threshold for a given propellant mass flow rate. The onset causes severe performance degradation and is marked by voltage fluctuations and increased electrode erosion. A couple theories have been proposed to explain the onset phenomena \cite{schrade1991}.\\

\textbf{Anode Starvation Theory}

According to this theory, the onset occurs when the current density at the anode reaches a level where there is not enough thermal energy to sustain the plasma discharge, which leads to the detachment of the plasma from the anode. The anode becomes 
"starved" of electrons, disrupting the thruster's stable operation.\\

The theory suggests that controlling the current density at the anode is crucial for preventing onset, and designs should focus on maintaining adequate thermal energy to sustain the plasma \cite{chanty1993}.\\

Recent findings have appeared to strongly support the explanation of onset phenomena through the Anode Starvation Theory. Notably, Professor Heiermann's numerical simulations at the University of Stuttgart affirm this effect. It shows that as the density of ions decreases in front of the anode, the resulting "starvation" leads to thruster instabilities under high electric currents. As a result, energy transfer becomes inefficient, and increased wear on the anode is observed, further disrupting stable operation\cite{heiermann2005}. 

\textbf{Full Ionization Theory}

This theory argues that onset occurs when the plasma in the thruster becomes fully ionized. When this occurs, the energy transferred to the plasma overheats the electrode walls, creating thruster instability. The theory emphasizes the importance of managing ionization levels within the plasma to prevent the excessive heating that leads to onset \cite{sanchez2014}. 

\section{Core Technologies}

\subsection{High Temperature Superconducting Coils}

Specific impulse and efficiency in AF-MPDTs increase with an increase in magnetic field strength \cite{lev2012}. However, achieving high magnetic field strengths with conventional electromagnets presents a challenge due to the significant electrical resistance. This is where the use of High-Temperature Superconductors (HTS) becomes appealing. HTS materials exhibit practically zero electrical resistance when operated below their critical temperature, enabling the creation of intense magnetic fields without significant energy losses. 

\begin{figure}[h]
    \centering
    \includegraphics[width=0.4\linewidth]{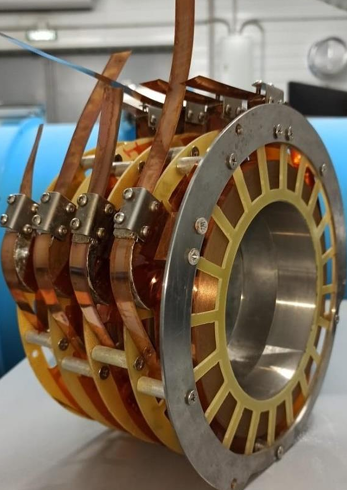}
    \caption{The four section HTS coil used in the SuperOx and MEPhI 25\,kW AF-MPDT \cite{voronov2020}.}
    \label{SuperOx-htsc}
\end{figure}

Recent experimental studies have shown that integrating HTS magnets into AF-MPDTs has led to notable improvements in thruster performance. The National Research Nuclear University (MEPhI) in Russia found that AF-MPDTs using HTS coils showed thrust increased up to 300\% and efficiency increased up to 700\% \cite{voronov2020}. Despite the success, several challenges remain with using HTS coils in AF-MPDTs. One of the most notable issues is the extreme temperature gradient within the thruster. The ions in the thruster can reach temperatures of over 29,000\,K \cite{myers19912}. At the same time, the HTS coils need to remain below temperatures as low as 60\,K \cite{bogel2022}. Hence, an extreme thermal barrier and an advanced and spaceflight-ready cryocooling system is required to maintain the superconductors below their critical temperature. Although cryocoolers like the Stirling and Reverse Turbo-Brayton (RTB) cryocoolers have flown in space, more research is required on their integration into AF-MPDTs. Overall, the integration of HTS coils into AF-MPDTs is promising, but more research and engineering development are needed to make it a spaceflight-ready technology at higher Technology Readiness Level (TRL). 

\subsection{Cryogenics}

The superconducting materials used in the electromagnet require cryogenic cooling systems to maintain the material under the necessary critical temperature. Although some HTS materials exhibit superconductivity at temperatures up to 77\,K, cooling to temperatures as low as 40\,K is often necessary. For AF-MPDTs between 5-20\,kW power class, the HTS coils require 10-15\,W of heat to be lifted by the cryocoolers \cite{bogel2022}. Hence, this is the range of cooling power necessary for the spaceflight cryocoolers to achieve. Historically, traditional cryocoolers, like the Stirling and RTB cryocoolers, have been used in space applications \cite{ross2003}. Researchers investigating superconducting AF-MPDTs are researching the integration of these cryocoolers into AF-MPDTs. \\

\subsubsection{Stirling Cryocoolers}

\begin{figure}[h]
    \centering
    \includegraphics[width=\linewidth]{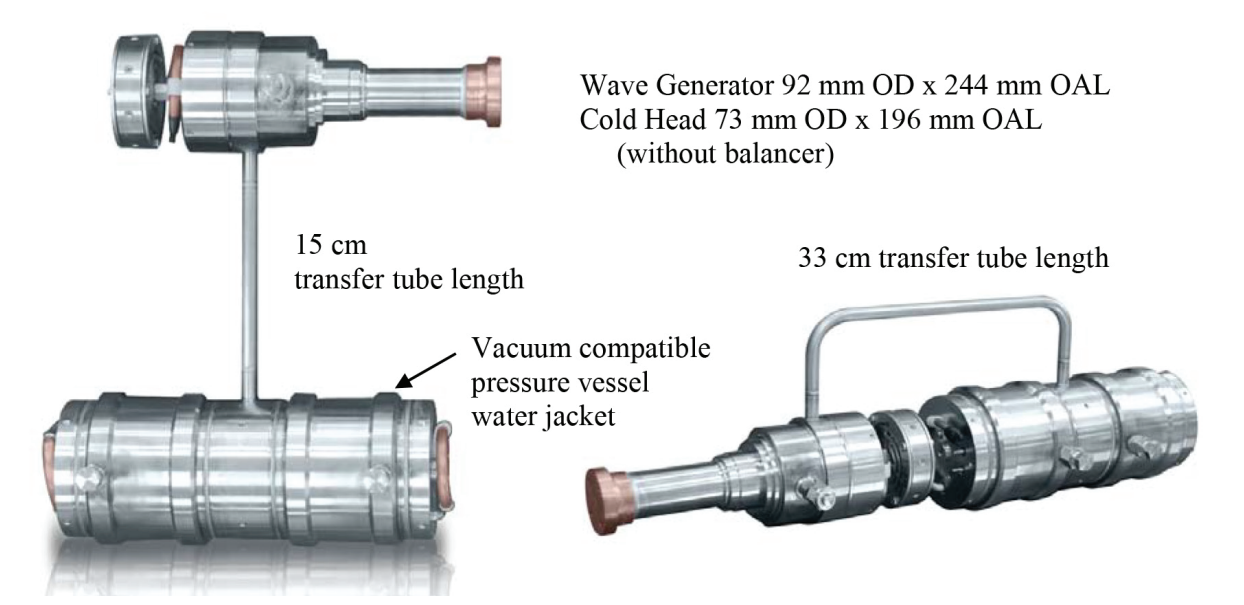}
    \caption{The Sunpower DS 30 CryoTel cryocooler with a high cooling capacity suitable for AF-MPDT heat loads \cite{iccsunpower}.}
    \label{ds30cryotel}
\end{figure}

Stirling cryocoolers are a favorite for cooling HTS coils due to their fast cool-down process, wide operating range, and high efficiency. Although there are downsides to the Stirling cryocooler, its outstanding cooling capabilities and relative manufacturing ease make it a great candidate for AF-MPDT applications \cite{xu2014}. Sunpower manufactures a `CryoTel' product line of cryocoolers, a Stirling cryocooler option for space operation \cite{sunpowercryotel}. The cryocooler weighs only 3.1\,kg with a cooling power of 10\,W to 25\,W. The operating temperatures range from 40\,K to 100\,K. Figure \ref{ds30cryotel} shows the Sunpower CryoTel DS 30 cryocooler which can reach a cooling power of 20\,W at a cold temperature of 60\,K \cite{iccsunpower}. \\

\subsubsection{Reverse Turbo-Brayton Cryocoolers}

Reverse Turbo-Brayton (RTB) cryocoolers have several unique advantages for space applications. RTB cryocoolers are efficient and have an extremely long lifespan. Additionally, they are relatively simple cryocoolers with few moving parts, allowing for better vibration control \cite{streit2013}. Creare, a world leader in RTB cryocoolers, has developed an RTB cryocooler suitable for space missions. Although it weighs 21\,kg, for an input power of 400\,W it produces a high cooling power of up to 12.6\,W. It has an operating range of 65\,K to 100\,K. This makes it a suitable use for cryocooling lower power class AF-MPDTs.

\subsection{Thermal Control}

Many companies have bespoke patented thermal management technologies with little literature available in the public domain, but much of these developments pertain to the material design of the thermal protection system and unique thermal properties of these materials being exploited in order to serve as effective insulation between the thruster and the HTS coil within the Applied-Field module. Thermal control systems in AF-MPDTs bear similarity to heat shield technologies used in atmospheric entry as well as to the thermal shielding applied in chemical rockets. 

\subsubsection{Passive Methods}  
Passive cooling methods are the most traditional approach for thermal control systems. They rely on taking advantage of the properties of materials to manage heat loads without active methods. The simplicity of passive methods makes them an attractive option for missions that want to minimize complexity. There are three main types of passive cooling methods which are commonly used in aerospace technology \cite{uyanna2020}.\\

\textbf{Heat Sinks}\\
Heat sinks are materials or structures that protect components by absorbing and storing heat. Modern heat sinks feature composite materials to minimize weight while enhancing heat absorption. They are generally effective for short-duration, high-heat flux scenarios. Metal based heat sinks have been commonly used in early designs, such as in the General Electric Mk II reentry vehicle \cite{glass2008}. Although heat sinks have had some success, testing has found a loss of strength leading to disintegration and melting. These failures led to the development of other methods for thermal protection systems \cite{allen1953}. \\

\textbf{Hot Structures}\\
Hot structures re-radiate absorbed heat using materials with high emissivity. These structures function as both a heat shield and a load-bearing structure, making them extremely useful for reusable launch vehicles. Current technology includes the use of metal matrix composites and ceramic matrix composites, which provide enhanced heat resistance and structural integrity. Research into ultra-high-temperature ceramics is currently ongoing to improve the thermal resistance of hot structures for the future \cite{tang2016}. \\

\textbf{Insulated Structures}\\
Insulated structures use multiple layers of thermal insulation to slow down heat transfer. These systems usually feature an outer radiative layer and an internal insulating layer. Materials like Reinforced Carbon-Carbon and silica-based insulation are generally used in insulated structures. Current research into nanomaterials and aerogels could lead to more effective and lighter insulative structures in the future \cite{li2019}. Moreover, Multi-Layer Insulation (MLI), layers of sheets typically made of aluminum foil, Mylar, or kapton, are used on spacecraft to insulate the full system as well as subsystems and components that need to be radiatively thermally shielded \cite{TIMMERHAUS200313}. Gaps are maintained between the layers which further contributes to the insulation the MLI blankets can provide \cite{BIRUR2003485}.\\ 

\subsubsection{Active Methods}
Active cooling methods are potentially the most sophisticated and effective way of managing extreme heat loads. Hence, it is generally used in AF-MPDT technology. The method involves using actively-moving coolants that absorb and remove the heat, which makes the system quite complex. There are two main active cooling methods used in thermal control systems \cite{uyanna2020}. \\

\textbf{Film Cooling}\\
Film cooling involves the injection of a coolant fluid along the vehicle's surface, creating a protective film that protects the structure from extreme heat. Film cooling is currently used in high-speed aircraft and rocket engines. Computer simulations and plasma wind tunnel tests can be used for research purposes to continue to improve the effectiveness of film cooling \cite{dellimore2010}. Research into the use of microjet arrays and advanced materials could enhance film cooling's effectiveness for future deep space missions \cite{sriram2009}.\\

\textbf{Transpiration Cooling}\\
Transpiration cooling utilizes a coolant that is passed through a porous surface, absorbing heat and creating a protective layer on the exterior of the vehicle \cite{bohrk2014}. This method is great at managing a large heat flux over extended periods. The German Space Agency has shown success with this method, achieving significant temperature reductions with minimal coolant flow \cite{vanforeest2009}. The development of more effective materials and coolants is key for advancing this technology. It can potentially be the most important method for AF-MPDT thermal control systems in the future. 

\subsection{Propellants}
The propellant choice for an AF-MPDT greatly impacts the performance of the thruster. Throughout the years, a wide range of propellants have been tested in an attempt to optimize the specific impulse and thrust efficiency across different power regimes\cite{kodys2005}.

\subsubsection{Lithium}
Lithium has been one of the most researched propellants for AF-MPDTs due to its low ionization potential, which minimizes ionization losses and improves thrust efficiency. Lithium AF-MPDTs usually operate at low voltages in the 250-500\,kW power range, with high efficiencies and moderate Isp around 3500-4000s\cite{kodys2005}. Lithium’s potential performance is significantly higher compared to other propellants, with efficiencies up to 69\% and specific impulses up to 5500s at 21\,kW. 

\subsubsection{Hydrogen}
Hydrogen is another propellant that has demonstrated high performance in AF-MPDTs, especially at lower power levels. Hydrogen thrusters show high efficiencies and specific impulses but generally operate at higher voltages compared to lithium\cite{kodys2005}. Research has indicated that hydrogen thrusters achieve higher efficiencies and specific impulses than argon thrusters under similar conditions \cite{myers1995}. Hydrogen thrusters at Tokyo University have achieved specific impulses of up to 6000 s at only 10\,kW, with efficiencies around 10-20\% lower than lithium thrusters at similar power levels \cite{sasoh1990}.

\subsubsection{Argon}
Argon has been another common choice for AF-MPDT propellants due to its availability and simplicity of handling. Argon typically results in lower performance metrics compared to lithium and hydrogen \cite{myers1991}. Efficiencies for argon propellants in the 10-100\,kW range are generally in the 5-25\% range, with specific impulses between 700-1800s \cite{sasoh1990}. 

However, radial segmentation of the mass flow rate between the cathode and anode has been found to enhance efficiency. By adjusting the distribution of argon flow, thermal load on the anode is reduced, which allows for significant energy savings. Experimental results show that a substantial portion of this saved energy is redirected to the engine's power, thereby improving thrust efficiency. This radial segmentation method has led to efficiency values close to 60\%, as energy that would have otherwise contributed to excess heating of the anode is instead utilized in propulsion. These findings suggest that, contrary to initial assumptions about argon's lower efficiency, the strategic flow segmentation can make argon a viable and efficient choice for AF-MPDTs \cite {heiermann2005}.

\subsubsection{Other Propellants}
Other propellants such as ammonia, nitrogen, helium, and cesium have also been explored, though they are less common. Each of these propellants presents unique advantages and challenges. For instance, ammonia has been successfully used in low-power AF-MPDTs for up to 100 hours, although significant erosion was observed. Cesium and helium have also been tested but are not as widely adopted due to various practical limitations.

\section{State of the Art: AF-MPDTs}

Current AF-MPDT research and development is focused on the use of HTS coils to generate the magnetic field. Generally, previously built AF-MPDTs have had a magnetic field strength of up to 0.4 Teslas \cite{zheng2021}. Creating a magnetic field with a magnetic field strength over 1 Tesla can theoretically maximize the performance of the thruster\cite{bogel2022}. However, with conventional electromagnets, that strength of a magnetic field requires bulky designs with high power usage. The high current densities of superconductors enable the creation of lighter, more compact, electromagnets, allowing the thruster to avoid the heavy mass and power consumption tolls of the conventional electromagnet\cite{collierwright2021}. Multiple entities around the world are currently engaged with superconducting AF-MPDT research, including groups in Russia, Germany, China, and New Zealand\cite{bogel2022}. The University of Stuttgart prototype currently is the only AF-MPDT design with a fully considered thermal management system; this remains a shortcoming of the Chinese, Russian, and New Zealand designs.

\subsection{New Zealand - Robinson Research Institute (RRI)}
In 2019, the Victoria University of Wellington and the RRI began pursuing the use of superconductors in MPDTs. They have completed several different computer models for various components of AF-MPDTs, most notably a 1 Tesla HTS Module\cite{olatunji2021}. They are currently focusing on developing the technology for use aboard small satellites and CubeSats and are planning a demonstration in 2025\cite{bogel2022}. 

\subsection{Germany - University of Stuttgart Institute of Space Systems (IRS)}
The Institute of Space Systems (IRS) is a research institute located at the University of Stuttgart in Germany. They have built a 100\,kW AF-MPDT and worked with Neutron Star Systems, a German space startup, to build a 5\,kW AF-MPDT, named SUPREME. Under the lead of the IRS, the SUPREME consortia is now developing both the 5\,kW AF-MPDT and the nominal 100\,kW AF-MPDT. The 5\,kW AF-MPDT features a light and compact superconductor and cryocooler system. They are currently researching coil configurations, cryogenic technologies, and magnetic field shapes \cite{bogel2022}. 

An important recent success comes from the Magnetohydrodynamic Enhanced Entry System for Space Transportation (MEESST) project, developed as part of a multi-institutional European constortium, including IRS, Neutron Star Systems, KU Leuven, and many other collaborators, which leverages HTS coils to generate strong magnetic fields that influence the plasma surrounding a spacecraft during atmospheric re-entry. This field pushes the ionized plasma layer away from the vehicle, significantly reducing the thermal load. By minimizing the need for traditional thermal protection systems, MEESST could lower spacecraft mass by up to 30\%, a critical improvement for cost-effective space missions. Additionally, initial testing has shown MEESST’s HTS-equipped plasma probe achieving up to an 80\% reduction in heat load, enabling a lighter and more compact approach to spacecraft shielding. By applying similar magnetic field configurations to AF-MPDTs, AF-MPDTs could achieve reduced thermal stresses on critical components, potentially increasing thrust efficiency and prolonging operational lifespan \cite{lani2023}.

\begin{figure} [h]
    \centering
    \includegraphics[width=\linewidth]{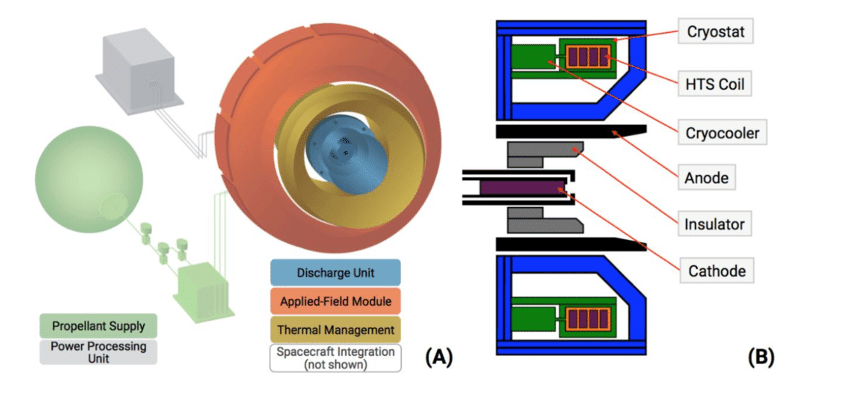}
    \caption{Schematic of the NSS SUPREME AF-MPDT \cite{collier-wright2019}.}
    \label{neutron}
\end{figure}

\subsection{Russia - SuperOx}
In 2020, Russia successfully utilized HTS coils in a 25\,kW AF-MPDT. The electromagnet produces a magnetic field with a strength of 1 Tesla with its 2G HTS SuperOx Tape. By introducing HTS coils into its thruster, the thruster reached a 60\% efficiency with 850\,mN of thrust and an ISP of 3840 seconds. This is a 300\% increase in thrust and specific impulse and a 700\% increase in overall efficiency\cite{voronov2020}. 

\begin{figure}[h]
    \centering
    \includegraphics[width=0.65\linewidth]{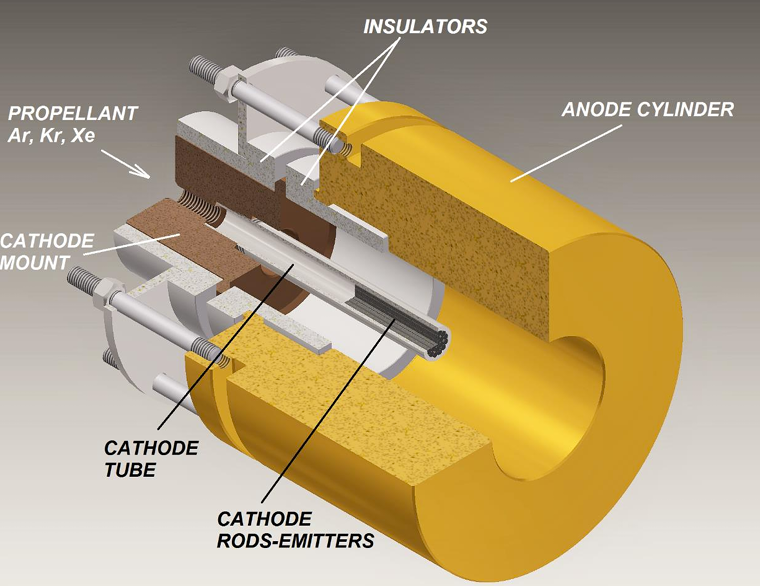}
    \caption{Design of the SuperOx AF-MPDT utilizing 2G HTSCs \cite{voronov2020}.}
    \label{russian}
\end{figure}

\subsection{China - Chinese Academy of Sciences (CAS)}
In 2022, the Chinese Academy of Sciences successfully designed and tested a 150\,kW AF-MPDT with superconducting coils. They reported an efficiency of 76\% and a specific impulse of over 5600 seconds. The thruster also demonstrated and confirmed the long-held theory that more powerful magnetic fields can reduce cathode erosion and offer a solution to AF-MPDT lifetime issues. However, it appears that this design did not incorporate a sufficient thermal management system. The lack of adequate cooling may have resulted in conditions where the cathode remained cooler than it would be in actual operating scenarios. In this cooler state, electron emission and plasma characteristics differ, resulting in data that likely underestimates the thermal stresses the system would encounter in actual operation. This discrepancy may lead to inaccurate assessments of the thruster's performance and efficiency \cite{zheng2021}\cite{bogel2022}. 

\begin{figure} [h]
    \centering
    \includegraphics[width=1\linewidth]{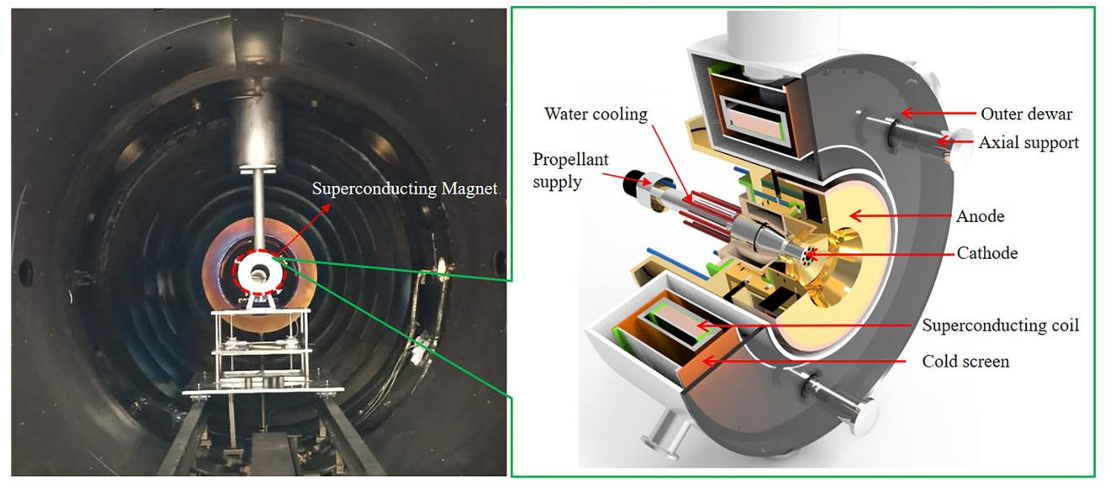}
    \caption{Design schematic of CAS150\,kW superconducting AF-MPDT \cite{zheng2021}.}
    \label{chinese}
\end{figure}

\section{Barriers to Spaceflight}

\subsection{Overview and Technology Readiness Level}

Many significant challenges remain before AF-MPDTs can be effectively utilized on spaceflight missions. The majority of the problems stem from the integration of the superconducting coils into the system, the cryocooling unit, the thermal management across the large gradient between the thruster and the cryocooler HTS coil, and the power supply required. Moreover, in order to reach TRL 9 (flown in space), the technology needs to be increased from TRL 4-5, which it is at present, up to TRL 8, where it would be ready for launch. In doing so, custom-designed testing components that mimic the function of each subsystem in order to perform individual component level testing, as well as the overall flight model subsystems being tested separately and then together while integrated, is required. This is a lengthy and technologically intensive process that requires years of design, building, testing and iterating with ample resources and use of plasma tunnels for the thruster testing.

\subsection{Spaceflight Cryocoolers}

\begin{figure}[h]
    \centering
    \includegraphics[width=\linewidth]{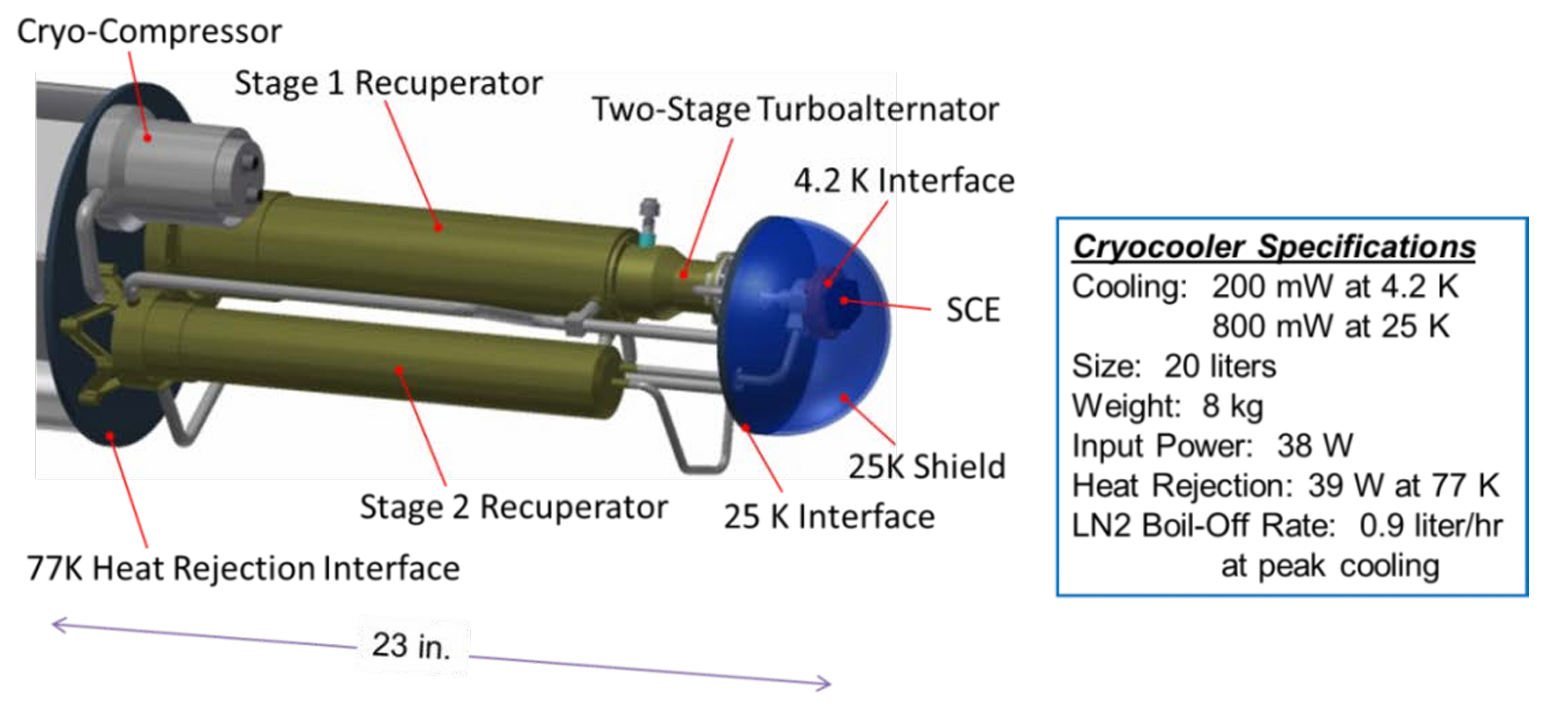}
    \caption{A schematic of a Creare Reverse Turbo-Brayton Cryocooler for low-temperature applications \cite{zagarola2017}.}
    \label{crearertb}
\end{figure}

A very high cooling power is needed to keep the superconducting coils under their critical temperature ($\sim$30-70\,K). Currently, no cryocooler with the required cooling power is compact and light enough to integrate into an AF-MPDT system. Potential solutions are currently being researched. High power Reverse Turbo-Brayton (RTB) cryocoolers are being developed by private companies around the world. For example, Figure \ref{crearertb} shows an RTB cryocooler developed by Creare that cools down to a temperature of 25\,K with a cooling power of 800\,mW \cite{zagarola2017}. These cryocoolers with designs that can reach much higher cooling powers ($\sim$2-10\,W) for higher power class AF-MPDTs ($\sim$5-10\,kW) are currently being developed but not yet established.

Moreover, solid state cryocooling solutions are being investigated for use in AF-MPDTs. However, it is clear that a closed cycle cryocooler with no open flow cryogen is needed for deep space missions, where the mass of the fluid tanks is significant and the fluid would run out. 

\subsection{HTS Developments}

One of the biggest challenges in implementing HTS technology into AF-MPDTs is the lifespan of the superconducting materials. Under the stresses of spaceflight, materials on spacecraft degrade over time due to solar radiation exposure, thermal cycling, and mechanical stress \cite{snov2019}. Of these considerations, the HTS coil will be subjected to thermal and mechanical, but unaffected by solar radiation given its shielding within the thruster unit. Research into improving HTS material durability and developing protective coatings or structures is necessary for deep space missions that could span multiple years. 

Although the REBCO superconducting material, the most widely used HTS in AF-MPDT development, is promising, more research is needed to further enhance its uses in space. REBCO has been used in spaceflight missions for HTS current leads; this is a fairly recent development for HTS coils that have been space-qualified \cite{Canavan_2024} where developments in establishing this as a space technology are ongoing \cite{Schlachter_2024}. Superconducting AF-MPDTs are generally considered to be at a TRL of 5 as numerous entities have proven its feasibility in a laboratory environment, but not yet in a space environment \cite{collier-wright2021}.

\subsection{Power Supply}

One of the biggest underlying problems slowing AF-MPDT development down for spaceflight missions is the immense power needed to efficiently run the AF-MPDT. Over 100\,kW of power is typically needed for high-efficiency, which is generally considered as a thrust efficiency of at least 30 percent \cite{jahn2002}. 

While in the past, high power AF-MPDTs were not feasible for deep space exploration due to those high power requirements, recent developments in spacecraft solar power technology have led to power projections of up to 200\,kW on future spacecraft missions\cite{brown2009}. 

However, the use of HTS material in the applied-field module creates more strain on the power supply. The cryocooler required to maintain the temperature of the HTS material adds to the AF-MPDT's power requirements. Many advanced power generation methods have been conceptualized for spaceflight, but not successfully conceived. For example, in 2005, NASA built a small nuclear reactor called Project Prometheus, which was expected to generate hundreds of kilowatts of power, but ultimately failed to do so \cite{patel2021}. 

\subsection{EP for Launch Vehicles}

Current MPDTs cannot be used for launch vehicles. Although MPDTs is almost ten times as propellant efficient as chemical propulsion, the mass flow rate is too low, which produces only a small thrust that is insufficient to reach the Earth's escape velocity \cite{heister2019}. AF-MPDTs are better suited for long duration missions in deep space, where their propellant efficiency and small thrust add up over time.

\section{Conclusions} 
In summary, AF-MPDTs are a high-potential developing technology that can greatly aid deep space exploration in the future. A majority of the recent developments in AF-MPDT technology have been geared towards the implementation of HTS coils, namely REBCO superconductors, into the applied-field module. Recent experiments and thruster prototypes have shown the current technology to be at TRL 4-5. In order to reach TRL 9, many problems stemming from the integration of HTS coils needs to be mitigated or solved, which could require years of design, building, and testing. The most pressing barriers to spaceflight are the management of the cryogenic cooling system needed to maintain the superconductivity in the coils and the development of power supplies in the range of hundreds of kilowatts. Both these systems need to be light and compact in order to be feasible for spaceflight applications. Over the next decade, AF-MPDTS are expected to steadily advance to higher TRL levels. If current development efforts persist, AF-MPDTS will become a viable propulsion option for deep space missions, offering a significant advantage in propulsion efficiency when compared to chemical propulsion.

\section*{Acknowledgements}
The authors would like to thank CCIR for funding support and the opportunity to conduct research on this topic. HR is supported by the Schmidt Science Fellows. The authors also thank Georg Herdrich of IRS Stuttgart for his helpful comments and discussion.

\bibliographystyle{elsarticle-num}

\bibliography{sample}

\end{document}